\documentclass[11pt]{article}

\begin{document}

\begin{titlepage}
 \
 \renewcommand{\thefootnote}{\fnsymbol{footnote}}
 \font\csc=cmcsc10 scaled\magstep1
{\baselineskip=14pt
 \rightline{
 \vbox{\hbox{hep-th/0409266}
       \hbox{UT-04-28}
       }}}

\baselineskip=20pt
\vskip 2cm

\begin{center}
{\bf \Large Modular Bootstrap of Boundary ${\cal N}=2$ Liouville Theory}\footnote{
Talk presented at strings 2004, June 28-July 2, Paris}
\end{center}

\bigskip

\vskip2cm

\begin{center}
 {\large Tohru Eguchi,\\
\vskip1mm
Department of Physics,\\
\vskip1mm
University of Tokyo, \\
\vskip1mm
Tokyo, Japan 113}
\end{center}

\vskip3cm

\begin{abstract}
We present our recent studies on the dynamics of boundary ${\cal N}=2$ Liouville theory. We use the representation theory of ${\cal N}=2$ superconformal algebra and the method of modular bootstrap to derive three classes of boundary states of the ${\cal N}=2$ Liouville theory. Class 1 and 2 branes are analogues of ZZ and FZZT branes of ${\cal N}=0,1$ Liouville theory while class 3 branes come from $U(1)$ degrees of freedom. We compare our results with those of  $SL(2;R)/U(1)$ supercoset which is known to be T-dual to ${\cal N}=2$ Liouville theory and describes the geometry of 2d black hole. We find good agreements with known results in $SL(2;R)/U(1)$ theory obtained by semi-classical analysis using DBI action. We also comment on the duality of ${\cal N}=2$ Liouville theory.
\end{abstract}

\vfill

\setcounter{footnote}{0}
\renewcommand{\thefootnote}{\arabic{footnote}}
\end{titlepage}

\newpage

\section{Introduction}
   
 Recently there has been a revival of interests in Liouville field theory,
due mainly to the reinterpretation of the old results of matrix model-Liouville theory    
as a typical example of gauge gravity correspondence \cite{MV,KMS,MTV,Martinec}. Matrix theory is 
interpreted as describing 
open string degrees of freedom living on D-branes while the Liouville field describes 
closed  string degrees of freedom.  
An exact map between the extended D-brane of Liouville theory
and macroscopic loop operator of matrix theory, for instance, has been discovered.
Analyses so far have been carried out using 
${\cal N}=0$
and ${\cal N}=1$ Liouville field theories. 

In this talk we would like to present  our recent studies of ${\cal N}=2$ Liouville theory based on the modular bootstrap approach \cite{ES1,ES2}. 
We recall that  ${\cal N}=2$ Liouville theory consists of a complex boson $\phi+iY$ and conjugate fermi fields
$\psi^+,\psi^-$. Here $\phi$ denotes the standard Liouville field coupled to the background charge ${\cal Q}$ and $Y$ is a compact boson which is not coupled to background charge.  
In our convention ${\cal N}=2$ Liouville system has a central charge
\begin{equation}
c=3+3{\cal Q}^2
\end{equation}
and the conformal dimension of a vertex operator $e^{\alpha\phi}$ is given by
${1\over 2}\alpha({\cal Q}-\alpha)$.

Interests in ${\cal N}=2$ Liouville theory 
 come from various directions:
 \begin{enumerate}
 \item
 Applications to string theory compactified on singular Calabi-Yau manifolds, NS5 branes etc..
 It is well-known that in the CHS background of the NS5 brane, transverse directions to the 5-brane are described by the Liouville field $\times$ $SU(2)$ WZW model\cite{OV,ABKS,GKP}. We then have
 \begin{equation}
 R^{5,1}\times R_{\phi}\times SU(2)_{\mbox{WZW}}\approx
 R^{5,1}\times \underbrace{R_{\phi}\times U(1)}_{\mbox{${\cal N}=2$ Liouville}}\times
 \underbrace{{SU(2)/U(1)}}_{\mbox{${\cal N}=2$ minimal}}
  \end{equation}
 where the first $U(1)$ factor is identified as the $Y$ field. Thus the space-time of NS5 brane is described by the 
 ${\cal N}=2$ Liouville field coupled to ${\cal N}=2$ minimal models. It is known that this geometry
 is T-dual to the ALE spaces and we will later compute the elliptic genus of ALE spaces by making use of the above description. 
 \vskip3mm
 \item
 It is known that ${\cal N}=2$ Liouville theory is T-dual to $SL(2;R)/U(1)$ super-
 coset theory \cite{FZZ2,HK} which describes the cigar geometry of 2-dimensional black hole
 \begin{equation}
 SL(2;R)/U(1) \mbox{ theory}=\mbox{T-dual of ${\cal N}=2$ Liouville}.
 \end{equation}
 Thus we can discuss 2d black holes by using ${\cal N}=2$ Liouville theory.
We will see below that unitary representations of $SL(2;R)/U(1)$ Kazama-Suzuki model are in fact identical to those of ${\cal N}=2$ Liouville theory \cite {DPL,ES2} and this gives a most straightforward proof of T-duality between these theories.   
\vskip3mm 
 \item
Bosonic sector of  ${\cal N}=2$ Liouville theory consists of fields $\phi$ and $Y$ and thus is identical to that of 
Liouville theory coupled to $c=1$ matter. Thus one expects a close relationship of ${\cal N}=2$ Liouville to bosonic models like sine-Liouville fields \cite{KKK}.
 \end{enumerate}
 
 \vskip2mm
 
 It is known that there are two different approaches to quantum Liouville theory;
 \begin{eqnarray}
 && \mbox{1. conformal bootstrap} ,\nonumber\\
&& \mbox{2. modular bootstrap}.\nonumber 
 \end{eqnarray}
 Conformal bootstrap \cite{ZZ,FZZ1,Teschner} is a detailed and complex analysis of the  Liouville system based on 
 conformal invariance and bootstrap in the presence of  world-sheet boundaries. It gives a detailed information on the dynamics of Liouville field and in particular  led to the discovery of boundary states and D-branes in
 Liouville theory. On the other hand, the modular bootstrap which is based on the representation theory and modular properties of character formulas  usually plays a secondary role in checking the consistency of the results of conformal bootstrap. 
 In ${\cal N}=2$ Liouville theory, however, it is difficult to carry out the conformal bootstrap approach due to technical complexity and we instead propose to use modular bootstrap in order to bypass the technical difficulties.

  Below we first discuss the representations of
${\cal N}=2$ Liouville  theory and then construct boundary states and their wave functions using of the modular properties of the character formulas. It turns out that there exist three different types of boundary states in ${\cal N}=2$ Liouville: class 1 branes
which are the analogues of ZZ branes in ${\cal N}=0$ theory \cite{ZZ}, class 2 branes which are the 
analogues of FZZT branes \cite{FZZ1,Teschner} and additional class 3 branes which come from the $U(1)$ degrees of freedom of ${\cal N}=2$ theory. 

We then use the relationship of ${\cal N}=2$ Liouville and $SL(2;R)/U(1)$ theory and  compare class 1,2,3 branes with the known D0,D1,D2 branes in the 2d black hole geometry \cite{RS}.  
 We find good agreements between the two except for a subtle discrepancy in the case of D1 brane. 
 
 In ${\cal N}=0$ and $1$ Liouville theories there exists the so-called duality symmetry of the theory where we exchange the parameter $b$ with $1/b$ where $b$ is related to the background charge ${\cal Q}$ as ${\cal Q}=b+1/b$. In the case of ${\cal N}=2$ theory an obvious duality symmetry is missing. Instead there exist two different types of Liouville potential terms
 \begin{eqnarray}
 &&\mbox{F term}:\hskip1cm \int d^2\theta e^{{1\over {\cal Q}}\Phi}+\int d^2\bar{\theta} e^{{1\over {\cal Q}}\bar{\Phi}},\label{F-term}\\
 &&\mbox{D term}:\hskip1cm  \int d^4\theta e^{{{\cal Q}\over 2}(\Phi+\bar{\Phi})}
 \label{D-term}\end{eqnarray}
where $\Phi$ denotes a chiral superfield with its lowest component $\phi+iY$. 
 These are both marginal operators preserving ${\cal N}=2$ superconformal symmetry and 
appear to play a dual role in the theory \cite{AKRS}. We will consider the issue of duality in ${\cal N}=2$ theory and point out that at a particular value of ${\cal Q}=1$ describing the singular K3 surface of $A_1$ type, this duality corresponds to the hyperK\"ahler rotation of the K3 surface.\\

\section{${\cal N}=0$ theory}

 Now let us first briefly recall the results of ${\cal N}=0$ Liouville theory and introduce the idea of modular bootstrap.  In ${\cal N}=0$ theory the central charge and the conformal dimension of a vertex operator 
 is given by (we use the convention $\alpha'=1$)
  \begin{eqnarray}
&& c=1+6{\cal Q}^2,\hskip3mm h[e^{2\alpha \phi}]=\alpha({\cal Q}-\alpha).
 \end{eqnarray}
When $\alpha={\cal Q}/2+ip$, the vertex operator $e^{2\alpha\phi}$ describes a primary field of continuous representation with momentum $p$ with dimension $h=p^2+{\cal Q}^2/4$.

In ${\cal N}=0$ theory there are basically two different representations;
\begin{enumerate}
\item
Identity representation
\begin{equation}
\chi_{h=0}(\tau)={1-q\over \eta(\tau)}\cdot q^{-{1\over 4}(b+{1\over b})^2}, \hskip3mm {\cal Q}=b+{1\over b}.
\end{equation}
\item
Continuous representations
\begin{equation}
\chi_p(\tau)={q^{p^2}\over \eta(\tau)}.
\end{equation}
\end{enumerate}
Identity representation has a singular vector at level $1$ while continuous representation is a non-degenerate 
representation. 
S-transform of these characters are given by
 \begin{eqnarray}
 &&\chi_{h=0}({-1\over \tau})=4\sqrt{2}\int_0^{\infty}dp \sinh(2\pi bp)\sinh({2\pi p\over b})\chi_{p}(\tau),\label{Strans-id}\\
 &&\chi_p({-1\over \tau})=2\sqrt{2}\int_0^{\infty}dp \cos(2\pi pp')\chi_p(\tau).
 \label{Strans-cont}\end{eqnarray}
 Now we introduce the boundary states $| ZZ\rangle$ and $| FZZT; p\rangle$ and identify the characters as expectation values of the evolution operator 
 \begin{eqnarray}
&& \chi_{h=0}({-1\over \tau})=\langle ZZ |e^{i\pi \tau H^{(c)} }| ZZ \rangle,\label{ZZbrane-rep}\\
 &&\chi_{p}({-1\over \tau})=\langle FZZT; p |e^{i\pi \tau H^{(c)}}| ZZ \rangle.
\label{FZZTbrane-rep}\end{eqnarray}
Here $| ZZ \rangle$ and $| FZZT; p\rangle$ denote ZZ and FZZT branes respectively and $H^{(c)}$ is the closed string Hamiltonian.
These boundary states are expanded in terms of Ishibashi states as
 \begin{eqnarray}
 &&| ZZ \rangle=\int_0^{\infty} dp' \Psi_0(p')|p' \rangle\rangle,\\
 &&  | FZZT; p \rangle=\int_0^{\infty} dp' \Psi_p(p')|p' \rangle\rangle 
 \end{eqnarray}
 where $|p' \rangle\rangle$'s are Ishibashi states with momentum $p'$  and $h={p'}^2+{\cal Q}^2/4$, and
 \begin{equation}
 \langle\langle p | e^{i\pi\tau H^{(c)}}| p' \rangle\rangle=\delta(p-p')\chi_p(\tau).
 \end{equation}
Here $\Psi_0(p')$ and $\Psi_p(p')$ are the disk one-point functions of ZZ and FZZT branes with a 
vertex operator $\exp{({\cal Q}/2+ip')\phi}$ insertion.

 In terms of these wave functions 
(\ref{ZZbrane-rep}),(\ref{FZZTbrane-rep}) are rewritten as
 \begin{eqnarray}
&&\chi_{h=0}({-1\over \tau})= \int_0^{\infty} dp |\Psi_0(p)|^2\chi_p(\tau),\\
&&  \chi_p({-1\over \tau})=\int_0^{\infty} dp' \,\Psi_0(p')\Psi_p^*(p')\chi_p'(\tau).
\end{eqnarray}
 By comparing with (\ref{Strans-id}),(\ref{Strans-cont})
 one finds
 \begin{eqnarray}
&& \Psi_0(p)\Psi(p)_0^*=4\sqrt{2}\sinh(2\pi  bp)\sinh({2\pi p\over b}),\\
 &&\Psi_p(p')^*\Psi_0(p')=2\sqrt{2}\cos(2\pi pp').
 \end{eqnarray}
Thus we can determine the wave functions 
\begin{eqnarray}
&&\Psi_0(p)=-2^{5/4}\cdot {2\pi ip {\hat{\mu}}^{ip\over b} \over \Gamma(1+i2pb)\Gamma(1+{i2p\over b})},\\
&&\Psi_p(p')=2^{1/4}{{\hat{\mu}}^{ip'\over b}\over 2\pi ip'}\Gamma(1-2ibp')\Gamma(1-{2ip'\over b})\cos(2\pi pp').
\end{eqnarray}
Here we have inserted some phase factors ($\hat{\mu}$ is related to the cosmological constant $\mu$ as $\hat{\mu}=\pi \mu \gamma(b^2)$ where $\gamma(x)=\Gamma(x)\Gamma(1-x)$).

This is the derivation of disk amplitudes using the modular bootstrap method: it is much simpler when compared with the  computations in  conformal bootstrap. Actually by the modular bootstrap method phase factors of the wave functions can not be determined, however, they cancel in the computation of cylinder amplitudes.  Thus using modular bootstrap method
 we can analyze the Cardy condition and  determine consistent boundary states. \\

Lessons we learn from  the above discussions are as follows:\\
\begin{enumerate}
\item
Identity representation does not appear in the closed string channel and closed strings states are spanned by the continuous representations $| p\rangle\rangle$.
Conformal dimensions of continuous representations are bounded from below
\begin{equation}
h(p)=p^2+{{\cal Q}^2\over 4}\ge {{\cal Q}^2\over 4}
\end{equation}
and thus the closed string sector has a gap in the spectrum.
This corresponds to the decoupling of gravity in the linear ditaton background.
\vskip3mm
\item
On the other hand, the identity representation does occur in the open string channel.
\vskip3mm
\item
One can check the consistency of the results of modular bootstrap with the conformal bootstrap analysis.
For instance, one can check the reflection property of the disk amplitude
\begin{equation}
\Psi_0(-p)=R(p)\Psi_0(p)
\end{equation}
where $R(p)$ denotes the reflection amplitude
 \begin{equation}
 R(p)=-{\hat{\mu}}^{-2ip/b}{\Gamma(1+2ip/b)\Gamma(1+2ipb)\over \Gamma(1-2ip/b)\Gamma(1-2ipb)}.
 \end{equation}
 
 \end{enumerate}
 
\section{${\cal N}=2$ theory}

Now we turn to the ${\cal N}=2$ theory. In the following we use the parametrization
\begin{equation}
\hat{c}={c\over 3}=1+{\cal Q}^2,\hskip3mm {\cal Q}^2={2K\over N}, \hskip2mm K,N \in {\bf Z}_{\ge 1}, 
\end{equation}
and the conformal dimension of a vertex operator is give by
\begin{equation}
h[e^{\alpha\Phi}]={p^2\over 2}+{{\cal Q}^2\over 8} \mbox{   for  } \alpha={{\cal Q}\over 2}+ip.
\end{equation}
(we use the convention $\alpha'=2$).
It is known that in ${\cal N}=2$ superconformal theory there exist three types of unitary representations, i.e. identity, continuous and discrete representations.
We denote their character formulas as $ch_{id}(\tau;z)$,$ch_{cont}(\tau;z)$ and $ch_{dis}(\tau;z)$, respectively. 
Angular variable $z$ is coupled to the $U(1)$ charge. 

 A continuous representation is a non-degenerate representation while the identity and discrete representation possesses a fermionic singular vector. Identity representation has also a bosonic singular vector. In the context of siring compactification we call them as graviton (identity), massive (continuous) and massless matter (discrete) representations.

 For the application to string compactification it is necessary to project the theory onto states with integral $U(1)_R$ charges. In order to ensure charge integrality condition we take the sum over spectral flows of irreducible characters and introduce the extended characters
 \begin{equation}
 \chi_*(\tau;z)=\sum_{n\in r+N{\bf Z}}q^{{\hat{c}\over 2}n^2}e^{2i\pi \hat{c}zn}ch_*(\tau;z+n\tau), \hskip2mm r\in {\bf Z}_N,\,\, *=id,\,cont,\,dis.
\end{equation}
 Note that we actually consider mod $N$ spectral flow to assure good modular properties of extended characters.
 Parameter $r$ runs over the range $0,1,\cdots,N-1$.

 Extended characters are parametrized as
 \begin{eqnarray}
 &&1. \mbox{ Identity representations}:  \nonumber \\
&&\chi_{id}(m;\tau);  \hskip1.5cm  m=2Kr,\hskip2mm  r\in {\bf Z}_N,\\
&& \hskip2cm 
  h={K\over N}r^2+|r|-{1\over 2},\hskip1mm Q={m\over N}+\mbox{sign}(r)\cdot 1, \,\, r\not =0 \\
 && \hskip2cm h=Q=0,\,\, r=0.  \nonumber \\
&& \nonumber \\ 
&& 2.\mbox{ Continuous representations}:  \nonumber \\
&&\chi_{cont}(p,m;\tau);  \hskip0.8cm p\ge 0, \hskip2mm m=2Kr,\,r\in {\bf Z}_{N}, \\
 &&  \hskip3cm h={p^2\over 2}+{m^2+K^2 \over 4NK},\hskip1mm Q={m\over N}. \\
&& \nonumber \\
&&3. \mbox{ Discrete representations}: \nonumber \\
&& \chi_{dis}(s,r;\tau); \hskip1cm m=s+2Kr,\, \, r \in {\bf Z}_N, \hskip1mm 0\le s\le N+2K-1,\\ 
&& \hskip1.2cm  h={Kr^2+(r+{1\over 2})s\over N},\hskip8mm Q={m\over N}, \hskip10mm 0\le r\le {N\over 2},\\
&&\hskip1.2cm   
 h={Kr^2+(r+{1\over 2})s\over N}-(r+{1\over 2}),\,Q={m\over N}-1,\, {-N\over 2}\le r \le -1.\nonumber 
\end{eqnarray}
 $h$ and $Q$ give the dimension and $U(1)$ charge of the representations. Note that the
quantum number $m$ is related to the $U(1)$ charge $Q$ as $Q=m/N$ (mod integer). Explicit form of these characters are given in  \cite{ES1}. In the following we consider the NS sector of the theory.
Characters in other sectors are obtained by half spectral flows. S-transformations of NS characters are given by \\

\noindent Identity representations:
\begin{eqnarray}
&&\chi_{id}(m;-\frac{1}{\tau})
= \nonumber \\
&& 
\hskip1cm \frac{1}{\sqrt{2NK}} \,\sum_{m'\in Z_{2NK}}\,
e^{-2\pi i \frac{mm'}{2KN}}\int_{0}^{\infty} dp' \, 
\frac{\sinh\left(\pi {\cal Q} p'\right)\sinh(2\pi \frac{p'}{\cal Q})}
{|\cosh \, \pi (\frac{p'}{\cal Q}+i\frac{m'}{2K})|^2}\, 
\chi_{cont}(p', m';\tau) 
\nonumber\\
&& \hskip1cm +\frac{2}{N}
\sum_{r'\in Z_{N}}\,\sum_{s'=K+1}^{N+K-1}\,
\sin(\frac{\pi (s'-K)}{N}) e^{-2\pi i \frac{m(s'+2Kr')}{2KN}}
\, \chi_{dis}(s',r';\tau)
\label{Liouville identity S}
\end{eqnarray}

\noindent Continuous representations:
\begin{eqnarray}
&& \chi_{cont}(p, m ; -\frac{1}{\tau})  
= \sqrt{\frac{2}{NK}}\,
 \sum_{m' \in Z_{2NK}}\, e^{-2\pi i \frac{mm'}{2NK}}  \int_{0}^{\infty}dp'\,
 \cos(2\pi p p') \,
\chi_{cont}(p', m' ; \tau) 
\label{Liouville massive S}\nonumber \\
&&
\end{eqnarray}

\noindent Discrete representations:
\begin{eqnarray}
&&\chi_{dis}(s,r;-\frac{1}{\tau})\nonumber \\
&&=\frac{1}{\sqrt{2NK}} \,\sum_{m'\in {\bf Z}_{2NK}}\,
e^{-2\pi i \frac{(s+2Kr)m'}{2NK}}e^{{i\pi m'\over 2K}}\int_{-\infty}^{\infty} dp'
{e^{-2\pi({s-K\over N}-{1\over 2}){p'\over {\cal Q}}}\over 2\cosh\pi({p'\over {\cal Q}}+{im'\over 2K})}
\, \chi_{cont}(p', m';\tau)
\nonumber\\
&&\nonumber \\
&&
\hskip1cm +\frac{i}{N}
\sum_{r'\in {\bf Z}_{N}}\,\sum_{s'=K+1}^{N+K-1}\,
e^{-2\pi i \frac{(s+2Kr)(s'+2Kr')-(s-K)(s'-K)}{2NK}}\, \chi_{dis}(s',r';\tau)\\
&&\nonumber  \\
&&
\hskip1cm +\frac{i}{2N}\,\sum_{r'\in {\bf Z}_{N}} 
\,e^{-2\pi i \frac{(s+2Kr)(s'+2Kr')}{2KN}}\,
\{\chi_{dis}(s'=K,r';\tau)-\chi_{dis}(s'=N+K,r';\tau)\} 
\label{Liouville massless S}\nonumber
\end{eqnarray}
Note that contributions from the "boundaries" of the range of  $s$, $K\le s\le K+N$ appear in the RHS of 
(\ref{Liouville massless S}).\\

The above transformation law has a peculiar structure, 
\begin{eqnarray}
&&\mbox{(continuous rep)}\stackrel{S}{\Longrightarrow}\mbox{(continuous rep)}\\
&&\mbox{(idenitity or discrete rep)}\stackrel{S}{\Longrightarrow} \mbox{(continuous rep)}+
\mbox{(discrete rep)}
\label{S of massless}\end{eqnarray}
Only a part of  discrete representations appear in the RHS (\ref{S of massless}).
Such a pattern of modular transformations was first observed in the representation theory of 
${\cal N}=4$ of superconformal algebra \cite{ET}.\\
 
 \begin{enumerate}
 \item
 We note that there appear no identity representations in the RHS of above formulas 
 (\ref{Liouville identity S}), (\ref{Liouville massive S}). (\ref{Liouville massless S}).
 \vskip5mm
 \item
 Only the discrete representations in the range $K\le s\le N+K$ appear in the RHS.
 \vskip4mm
 \item
 It is still possible to show that S-transformation squared equals $C$, $S^2=C$ where $C$ 
 is the charge conjugation operation \cite{IPT2}. This happens because the shift of the momentum contour becomes necessary in  the 2nd S-transform and from the fact that identity and discrete representations can be obtained from continuous representations at some complex values of the momentum.  
   
 \end{enumerate}
 
 \bigskip

Based on the above transformation law we propose that in the ${\cal N}=2$ theory closed string Ishibashi states are spanned by
\begin{equation}
\mbox{(continuous reps)+(discrete reps with }K\le s \le K+N).
\end{equation}
As we shall see below this spectrum agrees with those of the $SL(2;R)/U(1)$ coset theory.\\

We can construct class 1, 2, 3 boundary states corresponding to identity, continuous and discrete representations
using the modular transformation law.
In the case of class 1 and 2 branes, boundary  wave functions can be easily 
read off from the modular S matrices.
Class 1 boundary state (for the case $m=0$) is constructed as
\begin{equation}
 |B;id\rangle=\int_0^{\infty} dp'\sum_{m'\in {\bf Z}_{2NK}}\Psi_{id}(p',m')|p',m'\rangle\rangle+
 \sum_{r'\in {\bf Z}_{N}}\sum_{s'=K+1}^{N+K-1}C_{id}(s',r')|s',r'\rangle\rangle
\end{equation}
where
\begin{eqnarray}
&&\Psi_{id}(p',m')={1\over {\cal Q}}\cdot \left({2\over NK}\right)^{1/4}\cdot {\Gamma({1\over 2}+{m'\over 2K}+
i{p'\over {\cal Q}})\Gamma({1\over 2}-{m'\over 2K}+i{p'\over {\cal Q}})\over 
\Gamma(i{\cal Q}p')\Gamma(1+i{2p'\over {\cal Q}})},\\
&&C_{id}(s',r')=\left({2\over N}\right)^{1/2}\sqrt{\sin {\pi (s'-K)\over N}}
\end{eqnarray}
and $| p,m\rangle\rangle$ and $| r,s\rangle\rangle$ are Ishibashi states of continuous and discrete representations
\begin{eqnarray}
&&\langle\langle p,m |e^{i\pi \tau H^{(c)}}| p',m'\rangle\rangle=\delta(p-p')\delta_{m,m'}\chi_{cont}(p,m;\tau),
\\
&&\langle\langle s,r |e^{i\pi \tau H^{(c)}}| s',r'\rangle\rangle=\delta_{r,r'}\delta_{s,s'}\chi_{dis}(s,r;\tau).
\end{eqnarray}
We find that the amplitude $\langle B;id|\exp(i\pi \tau H^{(c)})|B;id\rangle$ reproduces the RHS 
of (\ref{Liouville identity S}).
Class 2 states are constructed in a similar manner.  By computing various cylinder amplitudes one can check
the Cardy consistency conditions.  

On the other hand, in the case of class 3 branes the situation becomes somewhat delicate due to the presence of
boundary terms in modular transformation (\ref{Liouville massless S}). In order  to cancel the effect of these terms   one has to consider a pair of discrete representations $[(r_1,s_1),(r_2,s_2)]$ with $(s_1+2r_1K)-(s_2+2r_2K)= {\bf N}\times \mbox{odd integer}$, and a combination of characters $\chi_{dis}(r_1,s_1;\tau)+\chi_{dis}(r_2,s_2;\tau)$. In this case Cardy condition are not necessarily  obeyed depending on the parameters $K,N$ of the theory.\\

\section{$SL(2;R)/U(1)$ theory}

It is well-known that
$SL(2;R)_k/U(1)$ supercoset theory is T-dual to the ${\cal N}=2$ Liouville. Central charges of these theories read as
\begin{equation}
\hat{c}=1+{2\over k}=1+{\cal Q}^2=1+{2K\over N}.
\end{equation}
Thus the level $k$ is related to the parameters $K,N$ by
\begin{equation}
k={N \over K}.
\end{equation}
By studying the representations of  $SL(2;R)/U(1)$ Kazama-Suzuki model, one finds that they 
are in fact in an one-to-one
correspondence with those of ${\cal N}=2$ Liouville and their character formulas are identical \cite{ES2}. 
(Relation of character formulas of bosonic $SL(2;R)/U(1)$ theory to those of ${\cal N}=2$ Liouville is given in \cite{IPT2}). Correspondence of parameters is given by
\begin{eqnarray}
&&\mbox{identity rep}:   \hskip1cm j=0,\\
&&\mbox{continuous rep}:   \hskip6mm j={1\over 2}+i{p\over {\cal Q}}, \\
&&\mbox{discrete rep}:    \hskip1cm  j={s\over 2K}
\end{eqnarray}
where $j$ denotes the spin of $SL(2;R)$ representation.

Path integral evaluation of toroidal partition function of $SL(2;R)/U(1)$ theory \cite{Gawedzki} 
shows that the theory contains continuous representations as well as discrete representations in the (improved) unitary range \cite{HPT,ES2,IKPT}
\begin{equation}
{1\over 2}\le j \le {k+1\over 2}.
\end{equation}
This agrees with the range $K\le s \le K+N$ of ${\cal N}=2$ Liouville under the correspondence $j=s/2K$.

 $SL(2;R)/U(1)$ theory describes the cigar geometry of 2d black hole and its boundary states are constructed by
 using Born-Infeld action and the results of $SL(2;R)$ theory \cite{RS}.
One can compare the results of ${\cal N}=2$ theory  with those of $SL(2;R)/U(1)$ \cite{IPT3,FNP}.\\

\hskip1.6cm Comparison of Boundary sates:

\vskip0.3cm

\hskip-1cm \begin{tabular}{|c|c|c|c|c|}
 \hline
\,\,\,Type of branes  \,\,\,&\,\,\,${\cal N}=2$ Liouville\,\,\, & & \,\,\,Type of branes \,\,\,& $SL(2;R)/U(1)$ \\ \hline
A &\mbox{class 1}  & & B &    D0 brane \\
 &\mbox{identity rep} & &  &     localized at tip of cigar\\ \hline
B &\mbox{class 2'} &  & A &       D1 brane \\
& \mbox{continuous rep} &&  &     \,\, extended along radial direction \,\,  \\ \hline
A & \mbox{class 3} & &  B &              D2 brane \\
& \mbox{discrete rep} &   &    &      wraps the whole cigar \\ \hline
A & \mbox{class 2} & &B &            D2 brane \\
 & \mbox{continuous rep} & &&     wraps a part of cigar \\  \hline
\end{tabular}

\bigskip
There exists an overall good agreement between the two. Wave functions of class 1 branes agree with those of 
$SL(2;R)/U(1)$ theory and also with the results of conformal bootstrap \cite{ASY1,ASY2}. 
We also have a match of class 3 and D2 branes with the identification
\begin{equation}
\sigma=\pi({s-1\over k}-{1\over 2})
\label{class3-D2}\end{equation}
where $\sigma$ is related to the gauge field strength on D2 brane \cite{RS}. Consistency of overlap of two D2 branes with $\sigma,\sigma'$ requires
\begin{equation}
\sigma-\sigma'=2\pi n{1\over k}
\end{equation}
where $n$ is an integer related to the D0 brane charges. This condition immediately follows from the above identification (\ref{class3-D2}).

However, in the case of class 2 branes
there is a delicate discrepancy with the D1 brane of $SL(2;R)/U(1)$ theory. Their wave functions have a form
\begin{eqnarray}
&&\Psi_{p,m}^{class2}(p',m')\approx f(p',m')\cos(2\pi pp')\\
&&\Psi_{p,m}^{D1}(p',m')\approx f(p',m'){e^{2\pi i pp'}+(-1)^{m'}e^{-2\pi i pp'} \over 2}.
\label{class 2'}\end{eqnarray}
Thus D1 brane has an extra phase factor $(-1)^{m'}$ as compared with the class 2 brane. Authors of \cite{FNP} call (\ref{class 2'}) as the class 2' brane.
 The extra phase factor is related to different reflection amplitudes for  A and B type states.
 Type B class 2 brane in fact does not have a good semi-classical limit unlike class 2' branes.
 It is suggested  that type A class 2 branes instead are well-defined and correspond to partially wrapped D2 branes in $SL(2;R)/U(1)$ theory \cite{FNP}.

\section{Singular Calabi-Yau manifolds}

When ${\cal N}=2$ Liouville theory is coupled to ${\cal N}=2$ minimal models, 
\begin{equation}
|\mbox{class 1 boundary state of Liouville} \rangle \otimes | \mbox{boundary states of minimal model}\rangle
\end{equation}
describe vanishing cycles of singular Calabi-Yau  manifolds: one recovers the correct intersection numbers by computing open string Witten index. See \cite{ES1} for details. 

One can also compute the elliptic genus of singular Calabi-Yau manifolds using ${\cal N}=2$ Liouville theory \cite{ES2}. Elliptic genus is defined by
\begin{equation}
Z(\tau;z)=Tr (-1)^{F_R}e^{2\pi i J_0z}q^{L_0-\hat{c}/8}\bar{q}^{\bar{L}_0-\hat{c}/8}
\end{equation}
and is invariant under smooth deformation of the parameters of the theory.
Here $F_R$ denotes the fermion number of the right-moving sector.
In the case of compact manifolds elliptic genus is known to have a good modular property and 
is a (quasi) Jacobi-form. It turns out that this is no longer the case in non-compact manifolds.

For instance, in the case of conifold one obtains the elliptic genus
\begin{equation}
{Z}_{conifold}(\tau,z)={1\over 2}{\theta_1(\tau;2z)\over \theta_1(\tau;z)}.
\end{equation}
In the case of ALE space with $A_{1}$ singularity it is given by
\begin{eqnarray}
&&{Z}_{ALE(A_{1})}(\tau,z)=-ch_0^{{\cal N}=4,\tilde{R}}(\ell=0;\tau;z) .
\end{eqnarray}
Here $ch_0^{{\cal N}=4}(\ell=0)$ denotes the massless  ${\cal N}=4$ character of isospin=$0$.
Elliptic genera of general $A_{n-1}$ type singularity are described in terms of the Appell function \cite{ES2}
\begin{equation}
{K}_{\ell}(\tau,\nu,\mu)=\sum_{m\in Z}{e^{i\pi m^2\ell\tau+2\pi im \ell\nu}\over 1-e^{2\pi i(\nu+\mu+m\tau)}}.
\end{equation}
Appell function is closely related to the character of discrete representations of ${\cal N}=2$ theory \cite{STT} and describe 
holomorphic sections of higher rank vector bundles on Riemann surfaces \cite{Pol}.

\section{${\cal Q}=1$ and Duality}

When the background charge takes a special value ${\cal Q}=1$, dimension of the target space $\hat{c}=1+{\cal Q}^2$ becomes $2$ and 
${\cal N}=2$ Liouville theory describes string compactification on
ALE space with $A_1$ singularity. Thus at ${\cal Q}=1$ ${\cal N}=2$ SUSY is enhanced to ${\cal N}=4$ .
Generators of ${\cal N}=2$ superconformal algebra (SCA) at ${\cal Q}=1$ are given by
\begin{eqnarray}
&&\hskip-8mm T=-{1\over 2}\left[(\partial Y)^2+(\partial \phi)^2
+\partial^2\phi+(\psi^{+}\partial \psi^{-}-\partial \psi^{+}\psi^{-})\right],\\
&&\hskip-8mm G^{+}={-1 \over \sqrt{2}}\psi^{+}(i\partial Y+ \partial \phi)- {1\over \sqrt{2}}\partial \psi^{+},
G^{-}={-1 \over \sqrt{2}}\psi^{-}(i\partial Y- \partial \phi)+ {1\over \sqrt{2}}\partial \psi^{-},\\
&&\hskip-8mm J_{U(1)}=\psi^{+}\psi^{-}-i\partial Y=i\partial H-i\partial Y
\end{eqnarray}
where we have bosonized the fermi fields  $\psi^+\psi^-=i\partial H$. $SU(2)$ currents of ${\cal N}=4$ SCA are given by 
\begin{eqnarray}
&&J^+_{\,SU(2)}=e^{iH-iY},\hskip3mm J^-_{\,SU(2)}=e^{-iH+iY},\hskip3mm 
J^3_{SU(2)}={i\over 2}(\partial H-\partial Y).
\end{eqnarray}
At ${\cal Q}=1$ Liouville potential terms (\ref{F-term}),(\ref{D-term}) become
\begin{eqnarray}
&&\hskip-1cm S_{+}=\int d^2 z \,e^{-\phi-iY-iH},\,S_3=\int d^2z \left(-i\partial Y-i\partial H\right)
(i\bar{\partial}Y+i\bar{\partial} H)e^{-\phi},\, S_{-}=\int d^2 z \,e^{-\phi+iY+iH}.\nonumber \\
&&
\end{eqnarray}
We find that $S_{\pm},S_3$ form a triplet under a new $SU(2)$ algebra $SU(2)'$ 
generated by
\begin{eqnarray}
&&J_{SU(2)'}^+=e^{iH+iY}, \hskip2mmJ_{SU(2)'}^-=e^{-iH-iY},\hskip2mm
 J_{SU(2)'}^3={1\over 2}(i\partial H+i\partial Y) .
\end{eqnarray}
$SU(2)'$ commutes with $SU(2)$ of ${\cal N}=4$ SCA 
and we have an $SU(2)\times SU(2)'$ structure

\newcommand{\mapright}[1]{%
 \smash{\mathop{%
  \hbox to 1.5cm{\rightarrowfill}}\limits^{#1}}}
\newcommand{\mapleft}[1]{%
 \smash{\mathop{%
   \hbox to 1.5cm{\leftarrowfill}}\limits^{#1}}}
\newcommand{\mapdown}[1]{\Big\downarrow
 \llap{$\vcenter{\hbox{$\scriptstyle#1\,$}}$ }}
\newcommand{\mapup}[1]{\Big\uparrow
 \rlap{$\vcenter{\hbox{$\scriptstyle#1$}}$ }}

\begin{equation}
\begin{array}{ccc}
G^+ \hskip-2mm=G^{+,+} & \mapright{J_{SU(2)'}^-}& G^{+,-} \\
 \mapdown{J_{SU(2)}^-} && \mapup{J_{SU(2)}^+}\\
 \stackrel{\null}{G^{-,+}} & \mapleft{J_{SU(2)'}^+ }& G^{-,-}\hskip-2mm =G^-
 \end{array}
 \end{equation}
 
 \begin{eqnarray}
&&G^{+,-}=+{1\over \sqrt{2}}e^{-iY}(\partial\phi-\partial H)+{1\over \sqrt{2}}\partial e^{-iY},\\
&&G^{-,+}=-{1\over \sqrt{2}}e^{-iY}(\partial\phi+\partial H)-{1\over \sqrt{2}}\partial e^{-iY}.
\end{eqnarray}
 
Action of $SU(2)'$ transforms $G^{\pm}$ into $G^{\mp}$ and  induces the change of complex structure. Thus it corresponds to the hyperK\"ahler rotations of K3 surface. 
Since $S_{\pm}$ an $S_3$ are transformed into each other under $SU(2)'$, the ${\cal N}=2$ duality amounts to a hyperK\"ahler rotation at ${\cal Q}=1$.
More details will be discussed in \cite{ES3}.\\

After this paper has been presented at strings2004 a new preprint appeared \cite{Hosomichi}
where new results of conformal bootstrap of ${\cal N}=2$ theory are reported.

\section*{Acknowledgements}

I would like to thank Y.Sugawara for his collaborations.


\begin{thebibliography}{100}

 \bibitem{MV}
J. McGreevy and H. Verlinde,
``Strings from tachyons: The c = 1 matrix reloaded,''
arXiv:hep-th/0304224.

\bibitem{KMS}
I. R. Klebanov, J. Maldacena and N. Seiberg,
``D-brane decay in two-dimensional string theory,''
JHEP {\bf 0307}, 045 (2003).


\bibitem{MTV}
McGreevy, J. Teschner and H. Verlinde, 
``Classical and quantum D-branes in 2D string theory,''
arXiv:hep-th/0305194.

 \bibitem{Martinec}
 E. J. Martinec,
 ``The annular report on non-critical string theory,''
arXiv:hep-th/0305148.

 \bibitem{ES1}
 T. Eguchi and Y. Sugawara,
 "Modular bootstrap for boundary N=2 Liouville theory"
 arXiv:hep-th/0311141.
 
 
 \bibitem{ES2}
 T. Eguch and Y. Sugawara,
 "$SL(2;R)/U(1)$ supercoset and elliptic genera of non-compact Calabi-Yau manifolds,"
 arXiv:hep-th/ 0403193.
 
 
 
 \bibitem{OV}
 H. Ooguri and C. Vafa,
 Nucl. Phys. B {\bf 463}, 55 (1996) [arXiv:hep-th/9511164].
 
 
 \bibitem{ABKS}
 O. Aharony, M. Berkooz, D. Kutasov and N. Seiberg, JHEP {\bf 9810}, 004 (1998) [arXiv:hep-th/9808149].
 
 \bibitem{GKP}
 A. Giveon, D. Kutasov and O. Pelc, JHEP {\bf 9910}, 035 (1999) [arXiv:hep-th/9907178].

\bibitem{FZZ2}
 V. Fatteev , A. B. Zamolodchikov and A. B. Zamolodchikov,
  unpublished.
  
    \bibitem{HK}
K. Hori and A. Kapstin,
JHEP {\bf 0108}, 045 (2001)
[arXiv:hep-th/0104202].

\bibitem{DPL}
L. Dixon, M. Peskin and J. Lykken,
Nucl. Phys. B {\bf 325}, 329 (1989).
 
  
 \bibitem{KKK}
V. Kazakov, I. Kostov and D. Kutasov 
 ``A matrix model for the two-dimensional black hole,''
Nucl.\ Phys.\ B {\bf 622}, 141 (2002)
[arXiv:hep-th/0101011].
 
 
  \bibitem{ZZ}
 A. B. Zamolodchikov and A. B. Zamolodchikov
 ``Liouville field theory on a pseudosphere,''
arXiv:hep-th/0101152. 
 \bibitem{FZZ1}
V. Fatteev , A. B. Zamolodchikov and A. B. Zamolodchikov,
 ``Boundary Liouville field theory. I: Boundary state and boundary  two-point function,''
arXiv:hep-th/0001012.
 \bibitem{Teschner}
 J. Teschner,
  ``Remarks on Liouville theory with boundary,''
arXiv:hep-th/0009138.  
 
 
 
 
 

 \bibitem{RS}
S. Ribault and V. Schomerus,
"Branes in the 2D black hole,"
arXiv:hep-th/031004.


\bibitem{AKRS} 
C. Ahn, C. Kim, C. Rim and M. Stanishkov,
``Duality in N = 2 super-Liouville theory,''
arXiv:hep-th/0210208. 
 
 \bibitem{ET}
 T. Eguchi and A. Taormina
  ``On The Unitary Representations Of  N=2 And N=4 Superconformal Algebras,''
Phys.\ Lett.\ B {\bf 210}, 125 (1988).
 
 
 \bibitem{IPT2}
 D. Israel, A. Pakman and J. Troost,
 "Extended $SL(2;R)/U(1)$ characters, or modular properties of a simple 
 non-rational conformal field theory,"
 arXiv:hep-th/0402085.
 
 \bibitem{Gawedzki}
 K. Gawedzki,
``Noncompact WZW conformal field theories,''
arXiv:hep-th/9110076.
 
 \bibitem{HPT}
 A. Hanany, N. Prezas and J. Troost,
 ``The partition function of the two-dimensional black hole conformal  field theory,''
arXiv:hep-th/0202129.
  
 \bibitem{IKPT}
  D. Israel, C. Kounas, A. Pakman and J. Troost, 
 "The partition function of the supersymmetric two-dimensional black hole and little string theory,"
 arXiv:hep-th/0403237.
 
 \bibitem{IPT3}
  D. Israel, A. Pakman and J. Troost, 
  "D-branes in N=2 Liouville theory and mirror,"
  arXiv:hep-th/0405259.
  
  
 \bibitem{FNP}
A. Fotopoulos, V. Niarchos and N. Perzas,
"D-branes and extended characters in $SL(2;R)/U(1)$,"
arXiv:hep-th/0406017.
 
 
 
 \bibitem{ASY1}
C. Ahn, M. Stanishkov and M. Yamamoto,
``One-point functions of N = 2 super-Liouville theory with boundary,''
arXiv:hep-th/0311169.

 
 \bibitem{ASY2}
C. Ahn, M. Stanishkov and M. Yamamoto,
"ZZ-Branes of N=2 Super-Liouville Theory,"
arXiv:hep-th/0405274.

 
 

 \bibitem{STT}
 A.   M. Semikhatov, A. Taormina and I. Y. Tipunin,
``Higher-Level Appell Functions, Modular Transformations, and Characters,''
arXiv:math.QA/0311314.


 \bibitem{Pol}
A. Polishchuk,
"M.P.Appell's function and vector bundles of rank 2 on ellipic curves,"
arXiv:math.AG/9810084.
 
 \bibitem{ES3}
 T. Eguchi and Y. Sugawara, in preparation.

 \bibitem{Hosomichi}
K.  Hosomichi,
"N=2 Liouville theory with boundary,"
arXiv:hep-th/0408172.
 
 
\end{thebibliography}
\end{document}